\documentclass[12pt]{article}
\usepackage[dvips]{color}
\usepackage{graphicx}
\usepackage{amsthm,amssymb,amsmath,graphics,times}

\newtheorem{theorem}{Theorem}
\newtheorem{lemma}{Lemma}

\begin{document}
\centerline{\Large \sffamily The Time Invariance Principle,
Ecological (Non)Chaos, and}

\centerline{\Large \sffamily A Fundamental Pitfall of Discrete
Modeling}

%\centerline{\Large \sffamily Time Invariance Principle and Why
%Logistic Map}

%\centerline{\Large \sffamily Does Not Model Any Process, Much Less
%Ecological Chaos}

%\centerline{\Large \sffamily The Time Invariance Principle and}

%\centerline{\Large \sffamily Why Ecologists Must Abandon Discrete
%Modeling}

\bigskip
\centerline{\sffamily Bo Deng\footnote{Department of Mathematics,
University of Nebraska-Lincoln, Lincoln, NE 68588. Email: {\tt
bdeng@math.unl.edu}}}

\bigskip
\noindent\textbf{Abstract: This paper is to show that most discrete
models used for population dynamics in ecology are inherently
pathological that their predications cannot be independently
verified by experiments because they violate a fundamental principle
of physics. The result is used to tackle an on-going controversy
regarding ecological chaos. Another implication of the result is
that all continuous dynamical systems must be modeled by
differential equations. As a result it suggests that researches
based on discrete modeling must be closely scrutinized and the
teaching of calculus and differential equations must be emphasized
for students of biology.}

% Time Invariance Principle, Experimental Reproducibility, Ecological Chaos,
% Ecological Stability, Discrete Modeling, Differential Equations,
% One-Life Rule, Logistic Growth, Beverton-Holt Model, Holling's Disc Function

\bigskip
\smallskip\noindent\textbf{1. Introduction.}
No models in ecology are better known than the Logistic Map, or have
played a greater role in the development of the chaos theory
(\cite{May74,Hass75,Hass76,Berr89,Loga92}). Surprisingly, however,
there is not a greater controversy than what was generated by the
model's prediction that one-species populations are inherently
chaotic.

The key prediction of the Logistic Map,
$x_{n+1}=Q(x_n,r):=rx_n(1-x_n)$, says that increase in the intrinsic
reproduction rate $r$ leads to chaotic dynamics for the population
$x_n$. Contradicting evidence existed before the chaos theory was
popularized in ecology. For example, in a 1971 study of an aquatic
system, McAllister and LeBrasseur (\cite{McAl71}) showed that
enrichment led to stable equilibrium. Extensive search for field
chaos came up equally empty-handed. For example, well-established
geographic patterns on microtine species (\cite{Hans91,Falc95})
showed that ecological systems tend to stabilize down the
north-to-south latitude gradient, correlating well with the ultimate
energy abundance from the Sun towards the equator. The most
comprehensive hunt for ecological chaos was down by Ellner and
Turchin (\cite{Elln95}). They used 3 different Lyapunove exponent
estimators on a large collection of empirical data. Out of their 21
field data sets, not a single set scored a positive Lyapunove
exponent by two of the 3 estimators. Out of their 20 lab data sets,
only two scored a positive Lyapunove exponent by two estimators. The
inescapable conclusion is overwhelming --- ecological chaos is not
to be expected in the wild. (Although laboratory chaos is possible
with stringent setups, such systems are never simple. In fact, the
dimension required is 3 or higher, c.f. \cite{Denn95,Cost97}.)

The glaring irreconcilability between the theory and reality can
only lead to one logical conclusion: the theory is wrong. Otherwise,
ecology would be the definitive branch of science that logic
imperative would have failed. The purpose of this paper is to make a
case that the Logistic Map and most discrete models used in ecology
and life sciences cannot be models for {\em any} physical process,
population dynamics in particular, and their predictions cannot be
{\em independently} verified by experiments.

\medskip\noindent\textbf{2. Time Invariance Principle.} This
conclusion rests on a fundamental principle of physics held since
the time of Copernicus in the 15th century that a physical law
should be the same anywhere and anytime in the universe. In other
words, a law must take the same mathematical form, derivable from
experiments carried out at independently chosen times and spaces. As
a result, the mathematical formulation of a law must be endowed with
such time invariance property. Taken to be self-evident, we state
the principle in the following formulation more suited for the
issues under consideration:

\begin{quote}
{\bf Time Invariance Principle (TIP):} {\em A physical law has the
same mathematical form to every independent choice of observation
time.}
\end{quote}

This principle has an important implication to dynamical systems as
laws of physical processes. To be precise, let  $x$ be the set of
state variables and $p$ be the set of parameters of a physical
process. As a dynamical system, $x$ changes in time $t$. Suppose an
observation is made at $t=0$ and the state is $x_0$. Another
observation is made at time $t>0$ and the state is $x_t$. Then, as a
physical law, $x_t$ is governed by a function, denoted by
$x_t=\phi_t(x_0,p)$, depending on the observation time $t$, the
initial state $x_0$, and the system parameter $p$. As a default
requirement, it must satisfy the unitary condition
\[
\phi_0(x_0,p)=x_0,
\]
that is, with time increment 0, the law $\phi_0$ leaves every state
fixed. Now by the Time Invariance Principle, if another observation
is made $s>0$ unit time later, the same function form
$(x_t)_{s}=\phi_{s}(x_t,p)$ must hold. Most importantly, the
function $\phi_t$ must satisfy the following group property and the
unitary condition
\begin{equation}\label{eqgroupproperty}
(x_t)_{s}=\phi_s(x_t,p)=\phi_s(\phi_t(x_0,p),p)=\phi_{s+t}(x_0,p)=x_{s+t},
\hbox{ and } \phi_0(x_0,p)=x_0,
\end{equation}
which together is referred to being {\it TIP-conforming}. That is,
if an observation is made $t$ time after the initial observation,
and another is made $s$ time later, then the result must be the same
if only one observation is made $s+t$ time after the initial
observation. More generally, the state at $s+t$ after an initial
$x_0$ is the same state at $s$ after an intermediate state $x_t$
which is the state at $t$ after the same initial $x_0$. A violation
of this property that $\phi_{s+t}(x_0,p)\ne\phi_s(\phi_t(x_0,p),p)$
implies that either such an ``experiment" is not reproducible, i.e.,
independent observation times inevitably lead to irreconcilable
conclusions, or such a functional form $\phi$ does not govern the
laws that the experiment is about to establish.

An immediate consequence to the Time Invariance Principle is the
following result.
\begin{lemma} If TIP-conforming dynamical system $\phi_t(x,p)$ is
continuously differentiable at $t=0$ and any $x$ in its domain of
definition, then $x(t)=\phi_t(x_0,p)$ must be the unique solution to
an initial value problem of a differential equation:
\[
%\left\{
%\begin{array}{l}
\displaystyle{\frac{dx(t)}{dt}}  = F_\phi(x(t),p),\ \ \ %\\
x(0)=x_0,
%\end{array}\right.
\]
where
\[
F_\phi(x,p)=\frac{\partial\phi_h}{\partial h}(x,p)\Big|_{h=0}
\]
is called the generating vector field of $\phi_t$.
\end{lemma}

\begin{proof}
Because $\phi$ is differentiable and is TIP-conforming
(\ref{eqgroupproperty}), we have the following derivative
\begin{equation*}
\begin{split}
\displaystyle{\frac{dx(t)}{dt}} & =\lim_{h\to
0}\frac{\phi_{t+h}(x_0,p)-\phi_t(x_0,p)}{h}=\lim_{h\to
0}\frac{\phi_{h}(\phi_{t}(x_0,p),p)-\phi_{0}(\phi_t(x_0,p))}{h} \\
& =\frac{\partial\phi_h}{\partial
h}(\phi_{t}(x_0,p),p)\Big|_{h=0}=F_\phi(x(t),p),\\
\end{split}
\end{equation*}
showing $x(t)$ is a solution of the equation. Since $F_\phi(x,p)$ is
continuous differentiable in $x$ because $\phi_t(x,p)$ is, the
solution to the initial value problem is unique.
%Interestingly, this argument shows that the Time
%Invariance Principle alone demands the tool of calculus.
\end{proof}

We now conclude that Logistic Map does not model any population
dynamics subject to time independent observations. More precisely,
we have the following result.
\begin{theorem} There does not exist a continuously differentiable,
TIP-conforming, 1-dimensional dynamical system $\phi_t(x,r)$ so that
$\phi_{t_0}(x,r)=Q(x,r)$ at any time $t_0$ and for all $x$ from any
interval containing $[0,1]$ for which $Q$ is the Logistic Map and
$r$ is the intrinsic growth rate with $r>3$.
\end{theorem}

\begin{proof}
By the preceding lemma, $x(t)=\phi_t(x_0,r)$ is the solution of an
autonomous differential equation $x'=F_\phi(x)$ generated by $\phi$.
Since the system is 1-dimensional, it does not allow periodic
solutions. However, the Logistic Map has a period-2 orbit for $r>3$
which would correspond to a periodic solution to the TIP-conforming
flow if it were true that $\phi_{t_0}(x,r)=Q(x,r)$ for some $t_0$. A
contradiction.
\end{proof}

This conclusion not only applies to the Logistic Map, but also to
most other discrete maps in ecology. Table \ref{tabmaps} lists some
popular discrete models in ecology. To be more precise, the same
argument can be used to show the following. The generalized
Beverton-Holt map is not TIP-conforming for $\gamma>1$ and large
$b$. The same for the Bernoulli model for $a>1$, the Richard map,
the Ricker map for large $r$. Applying the same argument for
2-dimensional TIP-conforming functionals shows they are solutions to
2-dimensional autonomous differential equations which do not allow
orientation reversing periodic orbits which occur to the
Nicholson-Bailey map.

For 3-dimensional or higher systems, the argument above for lower
dimensional systems do not apply. However, here is a diagnostic test
for possible TIP-nonconformity. More specifically, we certainly
assume that all biological processes are governed by physical laws
that are TIP-conforming, allowing time-independent observation and
verification on their states. Assume observation is made every unit
of time and $x_n$ is the state at time $t=n$. Because the state is
TIP-conforming, we must have
\[
x_n=\phi_1(x_{n-1},r)=\phi_1(\phi_1(x_{n-2},r),r)=\dots=\phi_1^n(x_0,r)
\]
where the exponent stands for iterative composition. Therefore,
\[
\phi_n(x_0,r)=\phi_1^n(x_0,r),
\]
that is, the $n$th iterative composition of $\phi_1$ must have the
{\em same functional form} as itself. This property can be used as a
diagnostic test for {\em probable} TIP-nonconformity. For example,
the $n$th iterate of the Logistic Map is a $2^n$-degree polynomial
with evolving coefficients for each $n$. This implies that the map
is very unlikely to be TIP-conforming because of the ever-changing
functional forms of its iterates or its TIP-conforming functional
would be extremely complex, in which case it is unlikely that such a
complex functional happens to satisfy a stringent condition such as
the TIP-conformity and at the same time arises from a relatively
simplistic modeling exercise that is typical of most discrete
modeling. This diagnostic test can be used to cast a serious doubt
on the TIP-conformity of the model under consideration. Unlike the
1- and 2-dimensional maps discussed above, for which the preceding
theorem provides a definitive means to determine their
TIP-nonconformity, we can only conjecture based on the preliminary
diagnostic test that the LPA map and the Leslie matrix are very
unlikely to be TIP-conforming. The same can be said for all
nonlinear models in cell-automata in games of life that they are
very unlikely to model any physical processes subject to
TIP-conformity. Without TIP-conformity, independent observations
cannot verify nor establish such maps as models, theories, or laws.

%%%%%%%%%%%%%%%%%%%%%%%%%%%%%%%%%%%%%%%%%%%%%%%%%%%%%%%%%%%%%%%%%%%%%%
%%%%%%%%%%%%%%%%%%%%%%%%%%%%%%%%%%%%%%%%%%%%%%%%%%%%%%%%%%%%%%%%%%%%%%
%%%%%%%%%%%%%%%%%%%%%%%%%%%%%%%%%%%%%%%%%%%%%%%%%%%%%%%%%%%%%%%%%%%%%%
\begin{table}%[ht]
\begin{center}
\caption{TIP-nonconforming Maps}\label{tabmaps}
\begin{tabular}{r|l}\hline
Generalized Beverton-Holt (\cite{Mayn73,Hass75,Hass76}) &
$N_{t+1}=\frac{bN_t}{1+(hN_t)^\gamma},\ \gamma\ne 1$
\\
Bernoulli & $N_{t+1}=aN_t (\textrm{mod}\ 1)$ \\
Logistic & $N_{t+1}=N_t[1+r(1-N_t/K)]$ \\
Richard (\cite{Rich59}) & $N_{t+1}=N_t\left[1+r(1-({N_t}/{K})^m\right],\  m\ne 1$ \\
Ricker (\cite{Rick54}) & $N_{t+1} = N_t\exp(r(1-N_t/K))$ \\
Nicholson-Bailey (\cite{Nich35}) & $ \left\{\begin{array}{l}
 N_{t+1} = N_t\exp(-aP_t) \\
 P_{t+1} = N_t(1-\exp(-aP_t))
\end{array}\right.
$ $\ ^{\ ^{\ ^{\ ^{\ ^{\ ^{\ ^{\ ^{\ ^{ }}}}}}}}}$
\\
LPA (\cite{Denn95}) & $\left\{\begin{array}{l}
 L_{t+1} = bA_t\exp(-c_{\rm el}L_t-c_{\rm ea}A_t)\\
 P_{t+1} = L_t(1-\mu_{\rm l})\\
 A_{t+1} = P_t\exp(-c_{\rm pa}A_t)+A_t{(1-\mu_{\rm a})}
\end{array}\right.
$ $\ ^{\ ^{\ ^{\ ^{\ ^{\ ^{\ ^{\ ^{\ ^{\ ^{\ ^{\ ^{\ ^{ }
}}}}}}}}}}}}$
\\
Leslie (\cite{Lesl45})  & ${\vec N}_{t+1}
=\left[\begin{array}{ccccc}
0 & f_1 & \cdots & f_{k-1} & f_k \\
s_0 & 0 & \cdots & 0 & 0 \\
0 & s_1 & \cdots & 0 & 0 \\
 & & \dots &  &  \\
 0 & 0 & \cdots & s_{k-1} & 0 \\
\end{array}
\right]^{\ ^{\ ^{\ ^{ }}}}\!\!\!\!\!\!\!{\vec N}_t$ \\
\hline

\end{tabular}
\end{center}
\end{table}
%%%%%%%%%%%%%%%%%%%%%%%%%%%%%%%%%%%%%%%%%%%%%%%%%%%%%%%%%%%%%%%%%%%%%%
%%%%%%%%%%%%%%%%%%%%%%%%%%%%%%%%%%%%%%%%%%%%%%%%%%%%%%%%%%%%%%%%%%%%%%
%%%%%%%%%%%%%%%%%%%%%%%%%%%%%%%%%%%%%%%%%%%%%%%%%%%%%%%%%%%%%%%%%%%%%%

\medskip\noindent\textbf{3. Discussions.}
The suggestion that biological research based on discrete maps is
build on a shaky scientific ground inevitably leads to a few
questions: One, is the Time Invariance Principle consistent with
other known physical principles? Two, what are the TIP-conforming
alternatives to discrete modeling in ecology? Three, can
TIP-nonconforming maps be justified and under what circumstances?
Four, contradicting to predictions by all discrete, chaotic maps in
theoretical ecology, why chaos is rare in the wild? We will exam
these issues for the remainder of the discussion.

\smallskip\noindent\textbf{Consistency With The Principles of
Relativity.} The Copernican idea that physical laws must be
universal in space and time has guided many great theories in
physics. Einstein's theories of special and general relativity are
two of the most celebrated examples. The Time Invariance Principle
is simply a corollary of the same idea that governs dynamical
processes.

One consequence of Einstein's theory of special relativity is that
there is no absolute time. TIP captures this time-relativity aspect
of his theory for the convenience of our discussion. As an example
to make the point, consider two inertial frames with Frame 2 moving
at a constant velocity $v$ with respect to Frame 1. Assume at the
origin of Frame 1 there is an on-going dynamical process. Let
$f_t(x)$ be the law deducted by observers of Frame 1 over a time
interval $t$ with initial state $x$. Due to the time dilation effect
of special relativity for Frame 2, observers in Frame 2 will not see
the same output $f_t(x)$ even though or precisely because they use
the same clock time interval $t$. Instead, they will see $f_\tau(x)$
for $\tau=t/\gamma>t$, $\gamma=\sqrt{1-v^2/c^2}$ with $c$ being the
speed of light. That is, both will see the same law but at two
different observation times because of their best intention to use
synchronized clocks at rest. This effect is equivalent to two
observers in the same inertial frame using independent sampling
times.

At the center of the special relativity lies the Lorentz
Transformation, relating space-time coordinates between two inertial
frames. Let $X=(t,x,y,z)$ be the space-time coordinate of Frame 1
and $X_v=(t_v,x_v,y_v,z_v)$ be the space-time coordinate of Frame 2
moving at a constant velocity $v$ with respect to Frame 1, say along
the same $x$-axis. Let $X_v=L_v(X)$ be the Lorentz Transformation
between the two coordinates. It is well-known that it satisfies this
self-consistent Compositional Invariance Property that
$L_u(X_v)=L_u(L_v(X))=L_w(X)$ where $w=(u+v)/(1+uv/c^2)$ with $c$
being the speed of light. That is, a third frame, Frame 3, moving at
a velocity $u$ relative to Frame 2 along the $x$-axis is a frame
moving at a velocity $w$ relative to Frame 1. The operation
$(u,v)\to w$ defines the so-called Lorentz group. The Lorentz
Transformation is one of the most well-known nontrivial and linear
maps in physics that is compositionally invariant. The
TIP-conforming group property is just a simpler kind of this more
generalized {\it Compositional Invariance Property}. Stochastic
matrixes (those which have non-negative entries and whose columns
each sums to 1) form another well-known class of compositionally
invariant linear maps.

\smallskip\noindent\textbf{Most Dynamical Systems Should Be Modeled By
Differential Equations.} It is a well-known simple fact that if
$\varphi(t,x_0,p)$ is the solution of a differential equation
$x'=F(x,p)$ with $t$ the time, $x_0$ the initial state at $t=0$, and
$p$ the parameter, then it always satisfies the TIP-conforming group
property
\[
\varphi(t+s,x_0,p)=\varphi(t,\varphi(s,x_0,p),p),
\]
for continuously differentiable right-hand side $F$.
%Interestingly, this argument shows that the Time
%Invariance Principle alone demands the tool of calculus.
Continuous-time and probabilistic processes can be modeled by
stochastic differential equations which also satisfy the
TIP-conforming group property (\ref{eqgroupproperty}), c.f.
\cite{LiLu05}. In additions, the same group property is satisfied
for hyperbolic PDE for age-structured populations, for parabolic PDE
for reaction-diffusion and traveling wave phenomena, and for delayed
differential equations. In such cases, the states lie in some
functional spaces which are infinitely dimensional. (Hence, by
differential equations for the remainder of the discussion we also
mean to include such infinite dimensional equations with or without
stochasticity, and as an extension by deterministic dynamical
systems we also mean to include probabilistic processes modeled by
stochastic differential equations for which it is the statistics in
the means, variances, distributions, etc. of some state variables
that become TIP-conforming, evolving deterministically.)

TIP-conforming processes are not restricted to deterministic
continuous-time processes only. In fact, true discrete probabilistic
processes can be TIP-conforming. For example, the process of coin
tossing has the same probability distribution satisfying
$\phi_n=\phi_1$ with $\phi_n$ representing the distribution at the
$n$th tossing, i.e., the same probabilistic law at any iterate of
the process. Also, stochastic processes modeled by Markov chains are
in general TIP-conforming because their transition matrixes have the
same functional form for all iterates.

The conclusion is that all (sufficiently differentiable,
continuous-time) TIP-conforming laws are governed by differential
equations. Hence, it is advisable to model all biological systems
whose states are subject to independent time observation by
differential equations in order to avoid the TIP-nonconforming trap
that discrete maps can easily fall into. For true discrete
probabilistic processes not modeled by differential equations, one
must check the models TIP-conformity, for which stochastic matrixes
represent one class of TIP-conforming, discrete, probabilistic
models.

The practitioner-dependent subjectiveness and arbitrariness of
picking time increments in discrete modeling cannot be more apparent
than modeling systems of various time scales. For example, when
modeling a system of two species, such as an algae-zooplankton
system or a plant-herbivore system, of which one operates at a
faster time scale and the other operates at a slower time scale,
what time step a discrete modeler should choose? If she picks the
fast time scale to be her discrete time increment, she will end up
fixing the slower species population as a parameter rather than an
evolving variable, and missing out the population's temporal booms
and busts. If she picks the slow time scale, she will end up
aggregating the dynamics of multiple generations of the faster
species, missing out its temporal booms and busts as well. Whatever
choice she makes, it is very likely that her choice will not be
honored by any independent modeler. Worse still, if her model is not
TIP-conforming, which is extremely likely, her model is doomed. On
the other hand, differential equation modeling does not have this
time-scale misalignment problem of discrete modeling, which is
easily dealt with by singularly perturbed differential equations,
which will be discussed further later.

\smallskip\noindent\textbf{TIP-Conformal Model --- Derivation by
One-Life Rule.} Any textbook derivation of the Logistic Map as an
one-species population model seems logically sound, yet it cannot be
substantiated by independent time observation. The inescapable
conclusion is that TIP must have been violated in all derivations.
Two alternative fixes are proposed below.

The first proposed fix follows the standard derivation of the
Logistic Map with modification to the functional form of the {\em
per-capita growth} of the species. The Logistic Map is the result of
assuming a linear functional for the per-capita growth as follows
\[
\frac{x_{n+1}-x_n}{x_n}=b-mx_n,
\]
where $b$ is the maximal per-capita growth rate, and $m$ is the
mortality rate due to intraspecific competition. All empirical data
(c.f. \cite{Odum71}), collected independently for different systems,
with uncoordinated time increments, point to a density-dependent
decreasing per-capita growth. That is, the decreasing monotonicity
is \textit{qualitatively} TIP-conforming for one-species per-capita
growth. Although the linear functional is qualitatively
TIP-conforming, it must have failed quantitatively, in particular at
high population density. For example, if the density is
$x_n=(9+b)/m$, then the per-capita growth is $b-m(9+b)/m=-9$,
implying paradoxically that each individual dies 9 times during the
given interval of time, possible only for mythological cat.
Furthermore, the per-capita growth any $k$th generation into the
future, $(x_{n+k}-x_n)/x_n$, fails to be strictly decreasing in
$x_n$ at hight growth rate $r$ for any $k\ge 2$. This implies that
the per-capita growth measured at a unit time interval $t=1$
decreases in the population density, but does not when measured at,
say, two units of time interval with $t=2$. In other words,
independent experiments would give contradicting outcomes in the
per-capita growth functionals if the Logistic Map were right, a
not-so-surprising paradoxical effect of the map's TIP-nonconformity.
Such inconsistencies are not limited to the Logistic Maps alone. In
fact, they plague all one-dimensional maps from Table \ref{tabmaps},
most of which are failed attempts to correct the Logistic Map.

Our proposed fix assumes instead
\[
\frac{x_{n+1}-x_n}{x_n}=\frac{b-mx_n}{1+mx_n}.
\]
Like the linear growth functional, it also decreases in $x_n$,
qualitatively TIP-conforming. More importantly, the per-capita
growth is bounded below from $-1$, and approaches $-1$ only as
$x_n\to\infty$. That is, {\it individuals die, but each dies at most
once in any fixed period of time }
--- a self-evident but both fundamental and universal principle for
all organisms which is referred to as the {\it One-Life Rule}. As a
result, the model is
\[
x_{n+1}=\frac{(b+1)x_n}{1+mx_n}:=\frac{rx_n}{1+mx_n}:=B_1(x_n,r,m),
\]
where $r=b+1>1$. This results in the Beverton-Holt model which was
first used by Beverton and Holt in 1956 for fishery studies
(\cite{Beve56}), but not derived from the One-Life Rule as we did
here.

It is simple to check that the Beverton-Holt model is TIP-conforming
--- the composition of two $B$ maps is another $B$ map:
\[
B_2(x_n,r,m):=B_1^2(x_n,r,m)=B_1(x_n,r^2,m(1+r)).
\]
In fact, one can demonstrate in general,
\[
B_k(x_n,r,m):=B_1^k(x_n,r,m)=B_1(x_n,r^k,m\frac{r^k-1}{r-1}),
\]
whose definition then can be extended to any continuous time $t$:
\[
x_t=B_t(x_0,r,m):=B_1(x_0,r^t,m\frac{r^t-1}{r-1})=\frac{r^tx_0}{1+m\frac{r^t-1}{r-1}x_0}.
\]
That is, any choice of fixed time increments leads to the same
functional form. More importantly, the discrete Beverton-Holt map is
simply the time-1 Poincar\'e map of the continuous counterpart, and
in this sense it cannot be regarded as a {\it true} discrete map.

Also, as a consistency check, one can easily show that the
per-capita growth over any time interval $t$ is always a decreasing
function of the initial population $x_0$ because
\[
\frac{x_t-x_0}{x_0}=
\frac{r^t-1-m\frac{r^t-1}{r-1}x_0}{1+m\frac{r^t-1}{r-1}x_0}.
\]
and that the One-life Rule is always satisfied since
\[
\lim_{x_0\to\infty}\frac{x_t-x_0}{x_0}=-1.
\]

The Ricker model is an improvement over the Logistic Map in that it
does obey the One-Life Rule. But it is not TIP-conforming and
produces the same pathological chaos prediction for one-species
population.

\smallskip\noindent\textbf{TIP-Conformal Model --- Derivation by Mass
Balance Law.} The second derivation can be best argued in terms of
stoichiometry. It recognizes that an organism is a package of
elemental elements, obeying the law of mass conservation. For
example, let us use carbon (C) as a basic unit to measure an
individual organism's biomass for a one-species system. Arbitrarily
fix a time increment, say $t=1$ for definitiveness. Let $x_n$ and
$x_{n+1}$ be the numbers of individuals for the current generation
and the ``next" generation respectively. Let $N$ be the amount of C
available in the interval, i.e., a constant flux in C. Let $c$ be
the amount of C that is needed during the period for each individual
which is to make to the next generation, i.e., the per-capita
maintenance cost in C. Let $a$ be the efficiency rate, which
measures the proportionality of the new generation that each
individual of the current generation gives rise to for each unit of
resource in C. It is the per-capita growth-to-consumption ratio.
Then $N-cx_{n+1}$ is the amount available for transition to the next
generation, and the product of $N-cx_{n+1}$, $a$, and $x_n$ gives
the next generation's population:
\[
x_{n+1}=(N-cx_{n+1})\times a\times x_{n}.
\]
Simplify to obtain
\[
x_{n+1}=\frac{rx_{n}}{1+mx_n}\ \ {\rm with}\ \ r=Na,\ m=ac,
\]
the same Beverton-Holt model obtained above.

The Beverton-Holt model for one-species population is qualitatively
consistent with all empirical studies cited in
\cite{Odum71,McAl71,Elln95}. In particular, any initial non-zero
population converges to an equilibrium:
\begin{equation*}
\begin{split}
\lim_{n\to\infty}x_{n} & =\lim_{n\to\infty}B_n(x_0,r,m)=\lim_{n\to\infty}B_1(x_0,r_n,m_n)\\
& =\lim_{n\to\infty}\frac{r^nx_0}{1+m\frac{r^n-1}{r-1}x_0}
=\frac{r-1}{m},
\end{split}
\end{equation*}
for which $r>1$ as a default assumption. In the context of the
stoichiometry for which $r=Na,\ m=ca$, we see that the greater the
nutrient influx $N$, the greater the stable equilibrium. The same
holds for smaller per-capita maintenance cost $c$ as well. The model
predicts that prosperity or efficiency or both promote stability,
not chaos, consistent with experimental findings such as
\cite{McAl71}.

\smallskip\noindent\textbf{TIP-Equivalence Example
--- The Logistic Equation.} As an illustration, consider the
continuous-time Beverton-Holt model
\[
x_t=B_t(x_0,r,m)=\frac{r^tx_0}{1+m\frac{r^t-1}{r-1}x_0}.
\]
Identical to the discrete case, it is also straightforward to check
the TIP-conforming group property
\[
B_{s+t}(x_0,r,m)=B_{s}(B_{t}(x_0,r,m),r,m).
\]
Thus, the generating differential equation to which $x_t$ is a
solution is obtained as
\begin{equation*}
\begin{split}
\frac{dx_t}{dt}&=\frac{d}{dh}B_{t+h}(x_0,r,m)|_{h=0}\\
 &=\frac{d}{dh}B_{h}(x_t,r,m)|_{h=0}
=\left(\ln r-m\frac{\ln r}{r-1}x_t\right)x_t,\\
\end{split}
\end{equation*}
the Logistic Differential Equation! The per-capital growth
\textit{rate} is linear which can be arbitrarily negative at high
population density. This does not violate the One-Life Rule which is
for the per-capital growth in a fixed time interval rather than the
instantaneous rate. The analysis above reaffirms a view that the
Logistic Differential Equation is a good population model. By our
argument it is because it is TIP-equivalent to the Beverton-Holt
model.

\smallskip\noindent\textbf{Justifications of TIP-nonconforming Maps
in Theoretical Ecology.} Continuous models of food chains of three
species or more can exhibit chaotic dynamics, which has been known
since the earlier days of the chaos theory, see for examples,
\cite{Hoge78,Gilp79,Hast91,McCa95,Smit94,Kuzn96}. One particularly
effective method to establish the existence of chaos in such models
is the method of singular perturbation, see for examples,
\cite{Deng99,Deng01,Deng02,Deng03,Deng04}. At the so-called singular
limit, some Poincar\'e return maps are one-dimensional, nonlinear,
and chaotic. However, such maps are not obtained by fixed time
interval samplings. Instead, they are event return maps. For
example, such a map may be defined when one of the predators reaches
local maximums in population density, for which to occur the moments
in time cannot be independently chosen. Such event Poincar\'e maps
are different from fixed time-step Poincar\'e maps in 3 critical
ways: First, the sampling time for the former covers a continuum
range of interval, conditioned on the occurrence of the event.
Second, the event Poincar\'e maps are not TIP-conforming, but the
fixed time-step Poincar\'e maps are like their continuous time
flows. Third, the fixed time-step Poincar\'e maps are always at
least 1-dimensional higher than their event counterparts.
One-dimensional event return maps are of many types, including the
unimodal type of which the Logistic Map is prototypical as
demonstrated in the cited references. In conclusion, discrete event
Poincar\'e maps are used as auxiliary means to understand continuous
models which generate them. They alone do not model the underlining
process in any time {\it independent} fashion.

Discrete maps also play an irreplaceable role in numerical
approximations. In fact, the Logistic Map is a discretization of the
Logistic Differential Equation. It is not TIP-conforming for any $r$
but it is a good approximation of the TIP-conforming continuous
model when $r$ is near 1. Similarly, TIP-nonconforming maps
generated as numerical schemes to approximate continuous models do
serve useful and important purposes in theoretical and practical
applications. But they are relevant only within their realistic
ranges. For example, it has always been an unjustifiable
extrapolation to large $r$ of the Logistic Map that it becomes
artificial and problematic. More {\it ad hoc} still than the
Logistic Map, the Ricker Map is not a discretization of any known
differential equation, further removed from TIP-conformity.
Similarly, a Leslie matrix should have been derived as a
discretization of the linearization of its PDE counterpart of the
age-structured population intended by the matrix, which
unfortunately was not always the case. Such a discretization of the
linear PDE should impose constraints on the time step and the age
increment. Outside such constraints, the Leslie model may become
problematic as is the case for the Logistic Map. With the above
rather thorough analysis of the Logistic Map, one implication seems
hard to miss that discrete models without underlining TIP-conforming
origins only have limited if not all questionable scientific values.

\smallskip\noindent\textbf{Enrichment and Efficiency Stabilization
Principles.} As pointed out in the introduction that empirical
studies do not support the hypothesis of chaotic one-species
population at high reproductive per-capita rate, or high efficiency
rate. In fact, the Beverton-Holt model implies diametrically the
opposite. For higher dimensional systems, the same opposing
dichotomy existed, and it can also be reconciled by TIP-conforming
models.

On one hand, chaotic dynamics do occur in models for three species
or more in food chains and webs
(\cite{Hoge78,Gilp79,Hast91,McCa95,Kuzn96,RaiV01,Deng01,Deng02,Deng03,Deng04,Bock04,Bock05,Deng06}),
as well as in laboratory models of 3-dimensional systems
(\cite{Cost97}). Also, almost all models exhibit the same
paradoxical effect that high reproductive efficiency leads to
chaotic population dynamics. Recent analysis from
\cite{Deng05,Deng06} concluded that this ``Chaos Paradox'' is an
artifact of the Malthusian exponential growth model for populations.
In addition to its projected unbounded growth fallacy, the
Malthusian model, $P'=rP$, also violates the One-Life Rule:
$(P(t)-P_0)/P_0=e^{rt}-1$, not going to $-1$ as $P_0\to\infty$. As
pointed out in \cite{Deng05,Deng06}, this Malthusian pathology hides
in almost all continuous models in the literature, and paradoxical
results are inevitable. The Enrichment Paradox (\cite{Rose71}), the
Biological Control Paradox (\cite{Luck90}), the Competition
Exclusion Principle (\cite{Arms80}), and the Chaos Paradox are of
the most notorious.

On the other hand, however, assuming logistic growth for all species
provides a sufficient remedy to all these paradoxes, the Chaos
Paradox in particular. For example, consider the following food
chain equations
\begin{equation}\label{dimensionalform}
\left\{\begin{split}
\dot X= & X\left(b_1-d_1-m_1X\right)-\frac{a_1X}{1+h_1a_1X}Y\\
\dot Y= & Y\left(\frac{b_2a_1X}{1+h_1a_1X}-d_2-m_2Y\right)-\frac{a_2Y}{1+h_2a_2Y}Z\\
\dot Z= & Z\left(\frac{b_3a_2Y}{1+h_2a_2Y}-d_3-m_3Z\right)
\end{split}\right.
\end{equation}
The result of \cite{Deng06} shows that all chaotic attractors
bifurcate into either limit cycles or steady states as the
top-predator's reproductive efficiency parameter $b_3$ increases,
and that the limit cycles further bifurcate into steady states if
the predator's reproductive efficiency parameter $b_2$ also
increases. In fact, one can show that all chaos attractors must
bifurcate into a steady state by increasing any two of the three
reproductive efficiency parameters in $b_1,b_2,b_3$. In other words,
efficiency promotes ecological stability, a result of evolution if
the hypothesis holds that evolution promotes species efficiency as a
survival fitness.

TIP-conformity is a necessary but not a sufficient requirement for
physical laws. Comparing to most studied in the literature, the food
chain model (\ref{dimensionalform}) above is significantly better
because of two features incorporated into the model. First, with the
inclusion of parameters $m_1,m_2,m_3$, intraspecific competitions
are taken into consideration for all species which leads to the
logistic growth rate for individual species (\cite{Verh38}) which in
turn conforms to the One-Life Rule. Second, the Holling Type II
predation functional form (\cite{Holl59}) is used for all predators.
The importance of this particular form lies in the fact that it is
TIP-conforming from its own derivation, and more generally it
satisfies the Composition Invariance Principle. To see this, we
recall from Holling's original mechanistic derivation that $r_e=aX$
is the number of prey, $X$, encountered in a unit time by one
predator, $Y$, or the encounter rate, and $h$ is the handling time
per kill of the prey, and $r=\frac{r_e}{1+hr_e}=\frac{aX}{1+haX}$ is
the number of handled kills in one unit time by one predator, or the
predation rate. This functional form is TIP-conforming for the
following reasons. If one breaks the handling time down to, say,
per-killing time $h_k$, and per-consuming time $h_c$ with
$h=h_k+h_c$, then Holling's derivation will give rise to the kill
rate $r_k=\frac{r_e}{1+h_kr_e}$ as a function of killing time and
encounter rate, and the consumption rate $r_c=\frac{r_k}{1+h_cr_k}$
as a function of consuming time and kill rate. It is straightforward
to check that
\[
r_c=\frac{r_k}{1+h_cr_k}=\frac{\frac{r_e}{1+h_kr_e}}{1+h_c\frac{r_e}{1+h_kr_e}}
=\frac{r_e}{1+(h_k+h_c)r_e}.
\]
That is, the rate function has the same mathematical form regardless
the temporal cut-off or definition of handling time, the essence of
TIP-conformity and compositional invariance.

In conclusion, theoretical and qualitative predictions of the food
chain model (\ref{dimensionalform}) are consistent with those of the
Beverton-Holt model as well as relevant empirical findings that
enrichment and efficiency promote ecological stability. The
consistency may not be coincidental, because all compartmental
constituents of the model are mechanistically TIP-conforming.

\smallskip\noindent\textbf{Concluding Remarks.}
Empirical data almost always are collected at discrete times.
Discrete modeling is an intuitive response to that reality to fit
discrete data by discrete models. However, a discrete model has
little to say about data collected at different discrete times of
the same process if the process permits. If it is not
TIP-conforming, it does not model the underlying process subject to
time independent observation. This may underlie many attempts via
stochastic inclusion to discrete modeling, attributing noise or
stochasticity as the chief cause of the irreconcilability between a
theory and reality when in fact TIP-nonconformity of the theory may
have been the problem. TIP-nonconforming event maps are secondary
structures of inherently higher dimensional TIP-conforming
differential equations. They rarely have a closed-form formula with
system parameters in plain sight for meaningful manipulations
because of the aggregating procedures that produce them. Even in
such cases, they are not closer in capturing the underlying physical
laws for the processes than the TIP-conforming differential
equations that model the processes. Given all these considerations,
this paper advocates a typical approach to use continuous models to
fit discrete data. Such models are open to the scrutiny of all
observations carried out at any discrete times. This approach makes
sure the models are necessarily consistent in its internal and
conceptual construct, allowing the modelers to modify and to refine
them within the realm of TIP-conformity.

Our TIP-equivalence result for fixed time-step Poincar\'e maps and
differential equations implies that 1 and 2 dimensional
TIP-conforming maps cannot be chaotic because 1 and 2 dimensional
differential equations of continuously differentiable vector fields
cannot be chaotic. Equivalently, chaotic 1 and 2 dimensional maps
must be TIP-nonconforming, and at the best arise as event Poincar\'e
maps of 3 or higher dimensional differential equations. As a result
such maps do not model any physical processes at a time-independent
fashion. Hence an ecological conclusion can be made unequivocally
that single- and two-species population dynamics cannot be chaotic.
That the controversy has lasted this long was due to the combination
of a few understandable factors. To name a few obvious: First, the
derivations of all popular discrete ecological models seemed
logical, but TIP-nonconforming nevertheless. Secondly, because of
their TIP-nonconformity, all predictions could not be independently
and objectively reproduced, leading to the inevitable confusing
state between a seeming reasonable theory and an uncompromising
reality. Thirdly, the field irreproducibility of all low dimensional
chaos theory was conveniently masked by the inherited
unpredictability of all chaotic systems. And fourthly, the
irreconcilability was also conveniently masked by a noisy reality
that is for most biological experiments and observations.

Comparing to differential equations, discrete maps are easier to
teach, easier to do research with. But we should not compromise the
Time Invariance Principle just for the simplicity appeal of discrete
modeling. TIP-conformity is the minimal necessary condition a
conceptually consistent model must satisfy. More importantly, the
requirement is fundamental to all branches of science, governing the
reproducibility of experiments. Because of these reasons, usage of
TIP-nonconforming maps is difficult to justify in most
circumstances. This conclusion has some important implications to
both research and training: Both past and future researches based on
discrete models must be scrutinized against their TIP-conformity and
be justified for their TIP-nonconformity. The subject of discrete
modeling may have to be de-emphasized in the classrooms and be
viewed through the lens of TIP-conformity. On the other hand,
training in calculus and differential equations must be further
enhanced and greatly emphasized for future generations of
theoretical biologists.

\bigskip

\noindent {\bf Acknowledgement:} The author gratefully acknowledges
comments and suggestions from his colleagues Glenn Ledder, David
Logan, Irakli Loladze from the Department of Mathematics, and Drew
Tyre, Svata Louda from the School of Biological Sciences at UNL.
Special thanks go to Dr. Jos\'e Cuesta of Universidad Carlos III de
Madrid, who helped the author to make the current argument against
Logistic Map's TIP-conformity precise and definitive.

\end{document}